\documentclass[a4paper]{article}

\usepackage{itgspeech2023}    
\usepackage{times}            
\usepackage[english]{babel}   
\usepackage[ansinew]{inputenc}
\usepackage[T1]{fontenc}      
\usepackage{graphicx}
\usepackage[colorlinks=false,pdfborder={0 0 0}]{hyperref}
\usepackage{units}
\usepackage[acronym, shortcuts, nohypertypes={acronym}]{glossaries}
\usepackage{booktabs}

\usepackage[natbib, bibencoding=utf8, citestyle=numeric, bibstyle=ieee, maxbibnames=999, maxcitenames=2, mincitenames=1, sortcites]{biblatex}
\usepackage[capitalise, nameinlink, noabbrev]{cleveref}

\nocite{*}
\bibliography{bibliography}

\newacronym{AI}{AI}{artificial intelligence}
\newacronym{ASR}{ASR}{automatic speech recognition}
\newacronym{CCC}{CCC}{concordance correlation coefficient}
\newacronym{CNN}{CNN}{convolutional neural network}
\newacronym{GMM}{GMM}{gaussian mixture models}
\newacronym{SVM}{SVM}{support vector machines}
\newacronym{LSTM}{LSTM}{long short term memory}
\newacronym{VAD}{VAD}{voice activity detection}
\newacronym{CE}{CE}{cross entropy}
\newacronym{UAR}{UAR}{unweighted average recall}
\newacronym{ACC}{ACC}{accuracy}
\newacronym{MAE}{MAE}{mean average error}
\newacronym{LDA}{LDA}{Linear Discriminative Analysis}
\newacronym{HMM}{HMM}{Hidden Markov Models}

\newcommand{\wtov}{wav2vec\,2.0}

\newcommand{\agender}{\mbox{aGender}}
\newcommand{\commonvoice}{\mbox{CommonVoice}}
\newcommand{\timit}{\mbox{Timit}}
\newcommand{\voxceleb}{\mbox{VoxCeleb2}}

\newcommand{\ie}{i.\,e., }

\newcommand{\cf}{{cf.\ }}

\title{Speech-based Age and Gender Prediction with Transformers}
\author{Felix Burkhardt$^1$, Johannes Wagner$^1$, Hagen Wierstorf$^1$, Florian Eyben$^1$, Bj\"orn Schuller,$^{1,2,3}$}
\address{$^1$audEERING GmbH, Germany,\\$^2$Chair EIHW, University of Augsburg, Germany, \\$^3$GLAM, Imperial College London, UK}

\begin{document}

\maketitle
\begin{abstract}
We report on the curation of several publicly available datasets for age and gender prediction. 
Furthermore, we present experiments to predict age and gender with models based on a pre-trained {\wtov}.
Depending on the dataset, 
we achieve 
an \acs{MAE} between $7.1$\,years and $10.8$\,years for age, 
and at least $91.1$\%\,\acs{ACC} for gender (\textit{female}, \textit{male}, \textit{child}).
Compared to a modelling approach built on hand-crafted features,
our proposed system
shows an improvement of $9$\% \acs{UAR} for age
and $4$\% \acs{UAR} for gender.
To make our findings reproducible,
we release the best performing model to the community as well as the sample lists of the data splits.
\end{abstract}



\section{Introduction}
\label{sec:intro}

The automatic detection of speaker age and gender has many use cases in human computer interaction, for example for 
dialogue 
adaption or market research. 
In contrast to subjective phenomena such as emotional arousal, the age of a person may be objectively determined, like for example body size, by an exact measurement. But, just like emotional arousal, age is only one of many factors that influence the acoustic speech signal~\citep{schoetz_2007}, and typically not to be predicted to the year.

We curated several publicly available datasets with respect to age labels and used them to train several age models based on a {\wtov} architecture.
We experiment on in- and cross-domain prediction, multi-head vs
single head models and the number of transformer layers to be used.
Finally we report the performance, using the 2010 paralinguistic challenge winner as a baseline.





Age and gender prediction based on machine learning as such has been investigated numerous times in the past decades.
A problem in this regard is the lack of benchmark datasets that could be used to compare approaches. 
There are several publicly available age annotated datasets like the SpeechDat II corpus\footnote{\url{https://tinyurl.com/speechdatii}},
{\commonvoice}~\cite{Ardila2020},
{\agender}~\cite{burkhardt-etal-2010-database},
{\timit}~\cite{timit},
{\voxceleb}~\cite{Nagrani2020},
or the NIST test set~\cite{Zazo2018},
but the studies we found used only some of these,
or not comparable train-development-test splits.
In addition,
the authors usually only report either regression
or classification results,
and use different metrics such as \ac{MAE}, \ac{ACC}, precision, recall, or \ac{UAR}.

\begin{figure}[h!]
    \includegraphics[width=.45\textwidth]{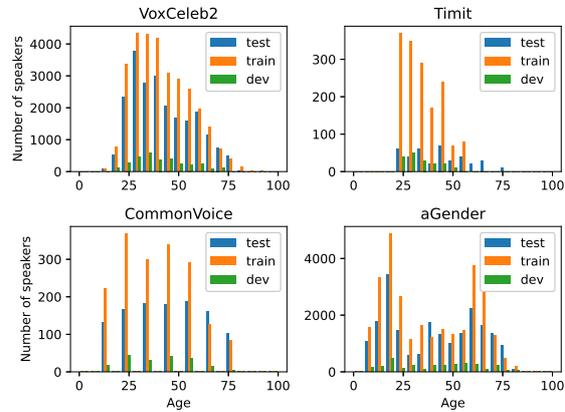}
    \caption{Distribution of speaker age (\#samples) in the datasets for the three splits ({\commonvoice} age in mid-decades). 
    }
  \label{fig:age_dist}
\end{figure}

During the 2010 Interspeech 
Paralinguistic Challenge \cite{Schuller2010}, age
classification was one of the topics.
\citet{Lingenfelser2010} report on the {\agender} dataset by fusing the results of ensemble classifiers trained on subgroups of a larger feature set and get 
 $42.4\%$\,\ac{UAR} on four age groups, the baseline being $46.2\%$\,\ac{UAR}.
\citet{Katerenchuk2017} use a similar configuration with respect to classifiers and feature sets to fuse acoustic and metadata for child speech detection.
Early studies are also based on the {\agender}~dataset~\cite{Bocklet2008, Feld2010}, both using \ac{GMM}/\ac{SVM} meta classifiers.
In \citet{Metze2007}, the best system  reaches an F1 value of $.54$ with a \ac{LDA} on \ac{HMM}s modelling phonemes on the SpeechDat II corpus. A human evaluation on a subset of the data reaches $.61$ F1 value. 
\citet{Sadjadi2016} describe
joint gender and age estimation using ivectors and Support Vector Regression, reaching 4.7 years \ac{MAE} on the NIST SRE 2010 telephony test set.
\citet{sanchez_2022} and \citet{tursunov2021age} both describe joint gender and age classification with a \ac{CNN} based on the {\commonvoice} dataset and report a recall of $76\%$ on six mixed gender-age groups and $74\%$\,\ac{UAR} on twelve mixed gender-age groups, respectively.
Comparing linear and logistic regression on ivectors with \ac{CNN}s using spectrograms, \citet{hechmi_2021} report $.98$ F1 for gender classification and $9.44$ years MAE for age regression. They use part of {\voxceleb} as a dataset with age ground truth labels estimated mainly from a Wikipedia lookup, and these labels are the basis for the {\voxceleb} data used in this paper, see Section \ref{sec:voxceleb} for details.
\citet{Zazo2018} propose a \ac{LSTM} recurrent network and report MAE of 6.58 years on the NIST test set.
\citet{Gupta2022} classified age using a \wtov{} model on the {\timit} dataset and report 5.54 years and 6.49 years \ac{MAE} for male and female speakers, respectively.
\citet{kwasny2020joint} reached a similar performance on {\timit} utilising a QuartzNet architecture, pre-trained on {\commonvoice} and {\voxceleb}, and fine-tuned on a joint age and sex prediction.
 With respect to gender prediction, authors usually refer to the biological sex. \citet{Levitan2016}  report on the {\agender} dataset three-class problem and reach 85\% accuracy on the test set with a Random Forest classifier and MFCC features.
 \citet{alnuaim2022speaker} use a pre-trained ResNet 50 and fine-tune it for gender on a balanced sub-set {\commonvoice}. 
 They report a recall of $.958$ on two gender groups
 on the test set of \commonvoice{}
 and a similar recall for cross-corpus performance.

With the paper at hand we see the following contributions:

\begin{itemize}
    \item We evaluate a novel system to estimate age and gender with fine tuned transformer models
    \item We present curated sample sets for train, development and test splits of publicly available data sets and make them available to the research community
    \item We compare a combined age and gender model to models specialized on a single task
    \item We present in-domain and cross-corpus results to examine the generalisability of the proposed system
    \item We investigate how many transformer layers are actually needed to properly model the tasks
    \item We release our best performing model to the public
\end{itemize}



\section{Datasets}\label{sec:database}
In order to test our approach on publicly available datasets, we considered the ones mentioned in the introductory Section \ref{sec:intro}. Most of them have drawbacks: The {\timit} and NIST datasets mainly contain young adults, Mozilla Common Voice is labelled with self-reported decades as age, and {\agender} is only available in $8$\,kHz telephone quality, {\voxceleb} might contain samples from the same speakers, but recorded in different years.
An overview on the age distributions per split can be seen in Figure \ref{fig:age_dist} and the numbers of samples and speakers per dataset and split in Table \ref{tab:dbs}.
All datasets are available for non-commercial research and the file lists can be accessed in the GitHub repository that accompanies this paper.\footnote{https://github.com/audeering/w2v2-age-gender-how-to}.

\subsection{\voxceleb}\label{sec:voxceleb}
\citet{hechmi_2021} report on a dataset which is based on a self-collected table for {\voxceleb} speakers. Because the authors (and the github repository) do not provide exact sample lists but only the speaker splits, we limited the number of samples per speaker to 20 samples. The original number of samples per speaker in the {\voxceleb} dataset has a mean value of 220.
We re-used the test set and split the train set randomly into 10\% development speakers and 90\% for training.
The age distribution for train and test splits is shown in Figure \ref{fig:age_dist}. As said, we followed the splits used in \citet{hechmi_2021} which have a rather large portion of test speakers. Although the age peak is still on the young side, it reflects the world age distribution better than, for example, {\timit}. 
As many samples are quite long ($>20$\,s), we used \ac{VAD} to segment the samples. 

\begin{table}[ht]
    \centering
    \footnotesize{
        \begin{tabular}{c | c c c} 
            Dataset & train & devel & test \\ [0.5ex] 
             \hline 
            {\voxceleb} & 30300 (1515) & 3120 (156) & 22100 (1105)\\ 
            {\commonvoice} & 1729 (118)& 186 (13)& 1110 (79)\\
            {\timit} & 1570 (157) & 170 (17) & 380 (38)\\
            {\agender} & 29553 (324) & 2974 (35) & 20549 (239) \\ [1ex] 
        \end{tabular}
    }
    \caption{Overview of the datasets: \#samples and \#speakers (in parenthesis).
    }
    \label{tab:dbs}
\end{table}

\subsection{\commonvoice}
The dataset is the result of a public data collection by the Mozilla foundation \cite{Ardila2020}. We used the \textit{de-validated} collection as a basis for our datasets.
Because the number of samples per speaker varies strongly (Mean: 55, STD: 296.6 
), we limited to 20 samples per speaker.
Because the age distribution is quite imbalanced, we furthermore tried to age-balance the test and training splits by selecting at most 20 speakers per age decade and gender, and then chose at most 7 speakers per age-gender group as a test set and the others for training.
As a development set, we randomly used 10\% of the training speakers.

\subsection{\timit}
The well known {\timit} datasets~\cite{timit} contains $16$\,kHz recordings of 630 speakers of eight major dialects of American English, each reading ten phonetically rich sentences.
Again, we tried to age-balance the test and training splits by 
first selecting at most 40 speakers per age decade and gender and then using 5 speakers per age-gender group as a test set, disregarding many of the speakers in their twenties. As a development set, we randomly used 10\% of the training speakers.

\subsection{\agender}
The {\agender} dataset~\cite{burkhardt-etal-2010-database} has been collected via telephone especially with age classification as a focus.
It 
has been used in the Interspeech 2010 Pralinguistic challenge\cite{Schuller2010}. The winning paper was \citet{kockmann10b_interspeech} and they reached $53.86$\,\ac{UAR} and $81.57$\,\ac{UAR} for the age (4 classes) and gender (3 classes) classification for the development set respectively with a meta classification based on \ac{SVM} and \ac{GMM}.
Because we wanted to be able to compare our results with the challenge, we used the official development set as our test set and took 10\% of the training speakers as a development set (the test set of the {\agender} dataset is secret and not accessible).
It is the only dataset that has been  collected solely for the purpose of biological age prediction and therefore does contain a substantial amount of children and already a balanced age structure.


\section{Architecture}\label{sec:architecure}

\begin{figure}[t]
    \centering
    \includegraphics[width=180pt]{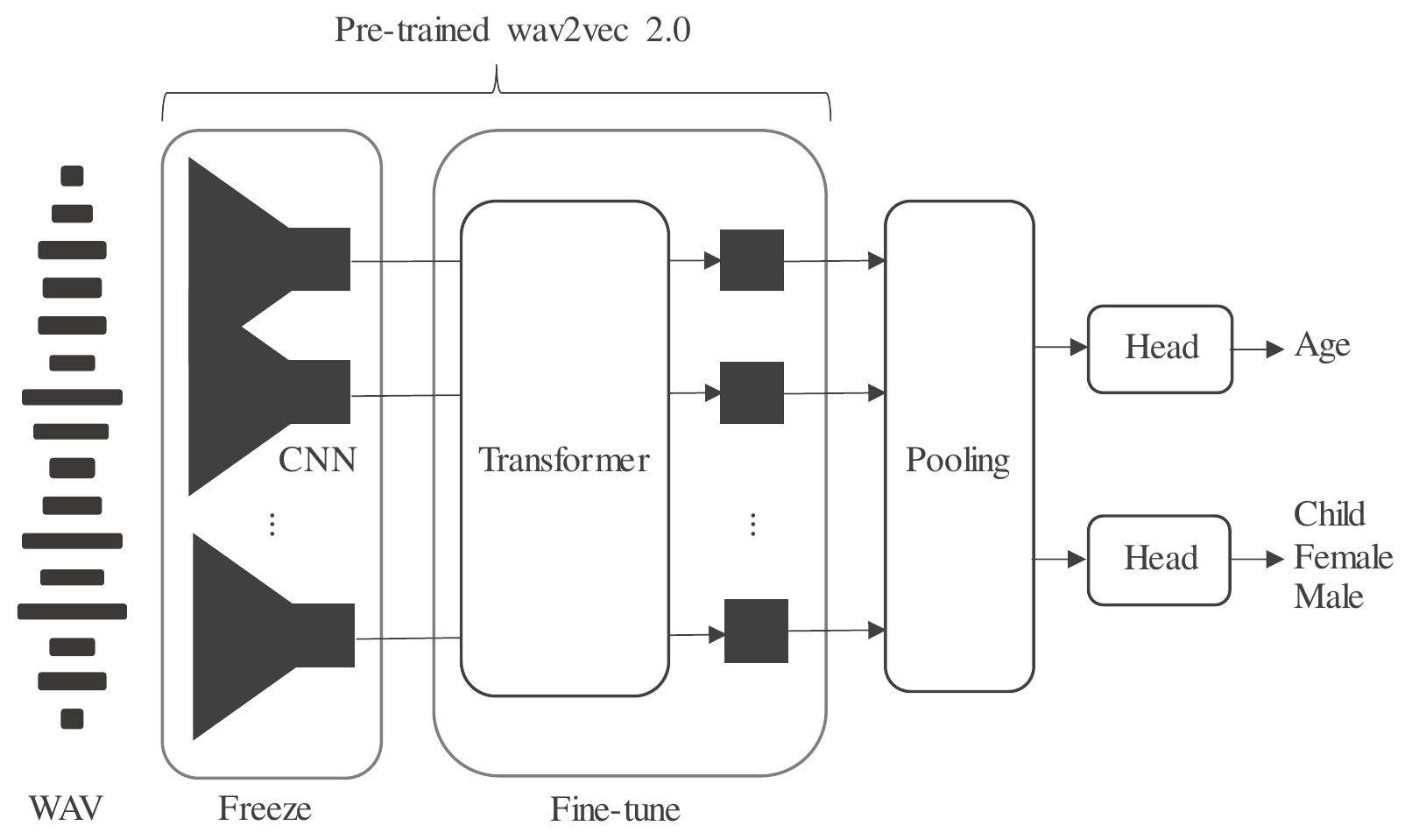}
    \caption{Proposed architecture built on {\wtov}.}
    \label{fig:architecture}
\end{figure}

Our proposed architecture is depicted in \cref{fig:architecture}.
It is built on {\wtov} \cite{baevski2020wav2vec}
with two custom heads to predict age and gender, 
respectively.
We do not start training from scratch,
but use the weights from a pre-trained model.
In our experiments we rely on
\emph{wav2vec2-large-robust}\footnote{https://huggingface.co/facebook/wav2vec2-large-robust},
a model pre-trained on read speech from 
Libri-Light (60k hours)
and {\commonvoice} (600 hours),
but also noisy telephone speech from
Fisher (2k hours)
and Switchboard (300 hours) \cite{hsu2021robust}.
We could show that
models fine-tuned on this variant
are generally more robust against noise
compared to models that have seen only clean speech
during the pre-training~\citep{wagner2023dawn}.

As input to the heads we use the pooled hidden states
(average pooling)
of the last transformer layer.
Each head consist of a fully connected layer of size 1024, a dropout layer,
and a final projection layer.
In case of age, we project to a single value predicting the \emph{age} in range 0 to 1, where 1 corresponds to a hundred years.
In case of gender, we project to three values expressing the confidence for being
\emph{child},
\emph{female},
and
\emph{male}.
During evaluation we decide in favor of the class with the highest value. 

For fine-tuning on the downstream task,
we use the ADAM optimiser with
a fixed learning rate of $1\mathrm{e}{-4}$.
Depending on the task we use two different loss functions:
\ac{CCC} loss for age, 
which we define as a regression problem;
\ac{CE} loss for gender,
which we define as a multi-class classification problem.
For backpropagation we use the average of the two losses.
We run for $5$ epochs with a batch size of $64$
and keep the checkpoint with best performance on the development set. 
As proposed in \citep{wagner2022dawn}
we freeze the \ac{CNN} layers but fine-tune the transformer ones. 
When using the term fine-tuning,
we will henceforth refer to this partial fine-tuning.
These models are trained using a single random seed,
for which the performance is reported.

In total, 
the model has $317.5$M parameters.
On a three second long input it performs $53.8$G MAC operations,
which took $34.2 \pm 5.2$\,ms when measured on a NVIDIA RTX A4000 (100 repetitions).
In \ref{sec:layers} we will discuss how these numbers can be reduced.
As learning framework we use PyTorch\footnote{https://pytorch.org/}
and rely heavily on the transformer library by HuggingFace \cite{wolf-etal-2020-transformers}.


\section{Experiments}\label{sec:experiments}

We will now report results
on predicting age and gender
of a speaker from her or his voice.
We treat gender detection as a classification task
and report results in terms of \acrfull{ACC} or \acrfull{UAR}
for the three classes child, female, and male.
In case of age,
which we model as a regression problem,
we report \acrfull{CCC}, and, for the {\agender} age groups: \acs{ACC} or \acs{UAR}.

\subsection{Single vs combined model}

\begin{figure}[t]
    \centering
    \includegraphics[width=\columnwidth]{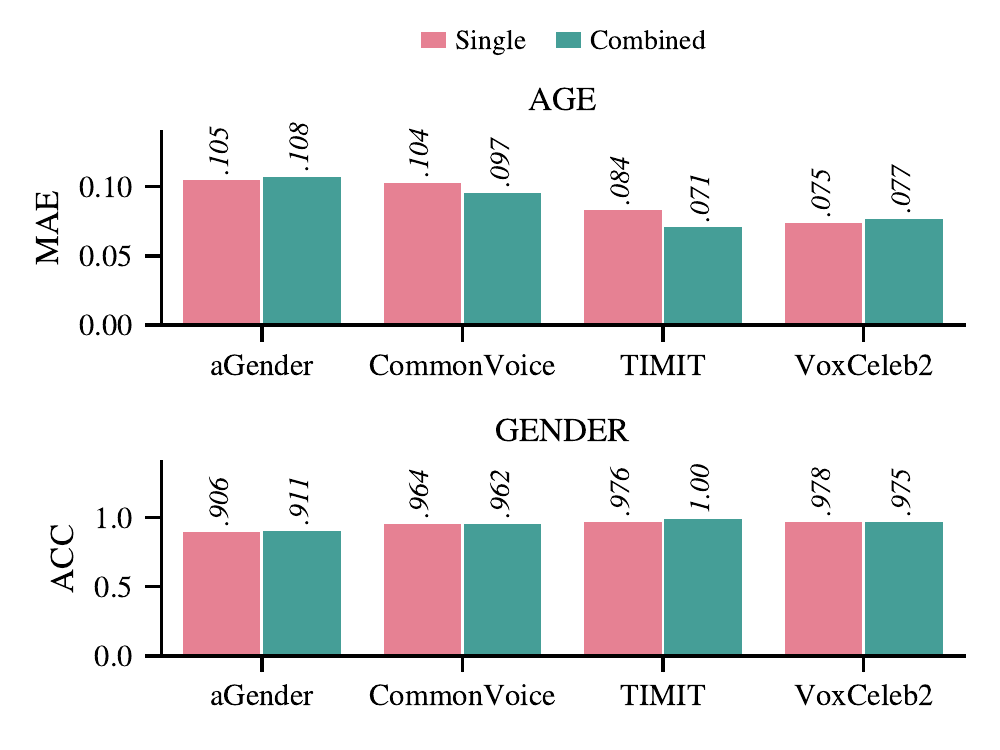}
    \caption{Performance of a model trained on a single task, \ie either age (MAE in years/100) or gender, and a combined model simultaneously trained on both entities. We see that accuracy stays more or less the same.}
    \label{fig:combined}
\end{figure}

We compare the performance of a 
\emph{combined} model trained on both tasks simultaneously
to that of models trained on a \emph{single} task,
\ie either age or gender.
For the latter,
we use the same architecture described in \cref{sec:architecure} 
but remove the other head.

Results are summarised in \cref{fig:combined}
and it is not difficult to recognise that the 
single and combined model perform almost identical.
We can conclude that,
although the combined model does not benefit from the information of the other channel,
it is well able to learn both tasks at once.
Since this (almost) halves the resources needed to run two separate models,
a combined model should be preferred in a multi-task setup.
Throughout the remaining of the paper,
we will report results for the combined architecture
without explicit mention.

\subsection{Cross-corpus evaluation}

\begin{figure}[ht]
    \centering
    
    \includegraphics[width=\columnwidth]{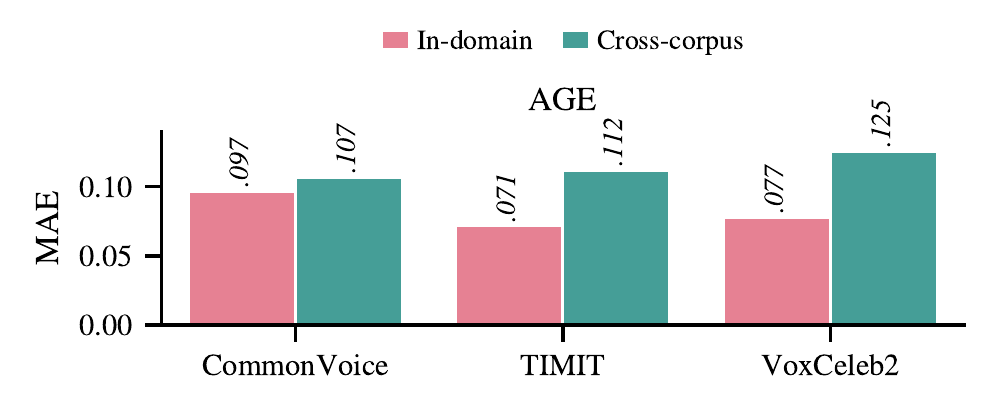}
    
    \includegraphics[width=0.9\columnwidth]{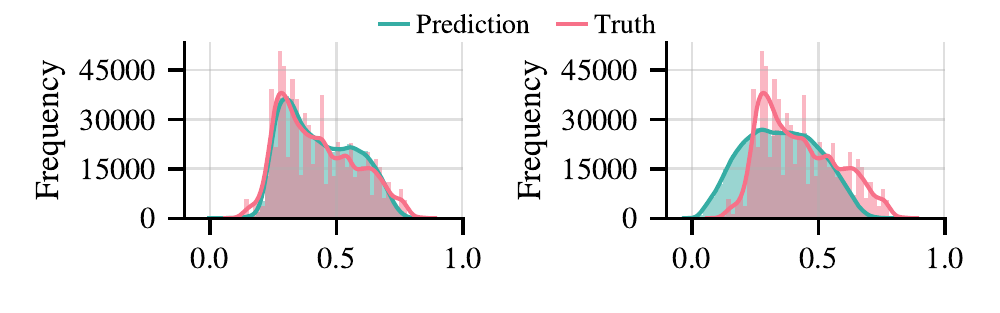}
    
    \includegraphics[width=\columnwidth]{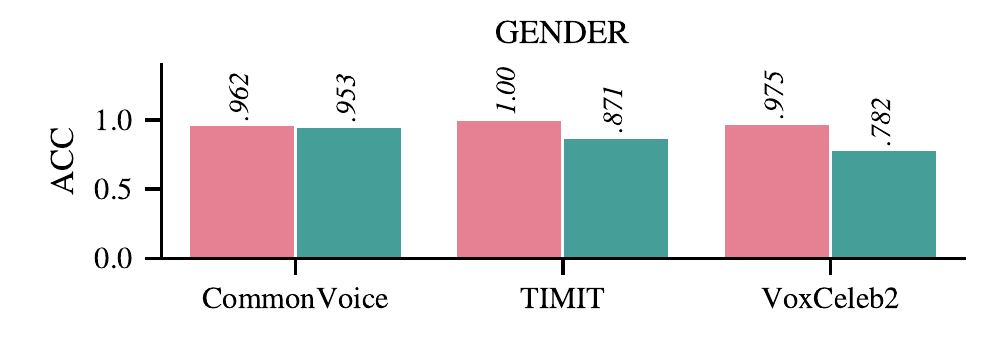}

    \vspace{-0.3cm}
    
    \includegraphics[width=0.9\columnwidth]{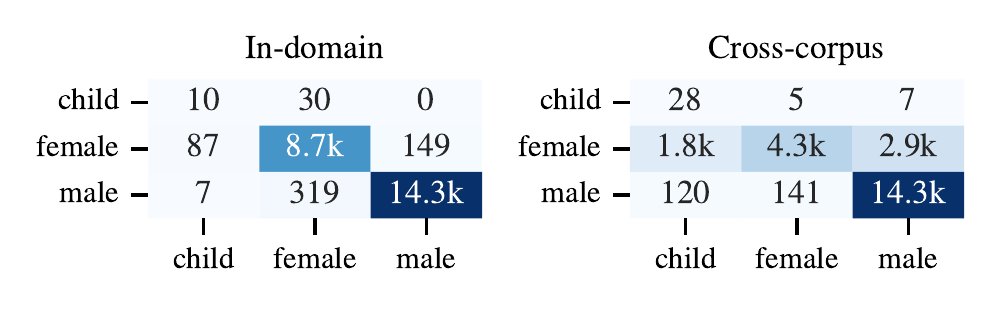}    
    
    \caption{In the cross-corpus condition, the model has not seen data from the dataset it is evaluated on. Results are compared with a model trained in an in-domain fashion. The cross-domain model performs poorly in detecting females (\cf confusion matrices) and predicts speakers younger than they are (\cf distribution plots) (age: MAE in years/100).}
    \label{fig:cross-corpus}
\end{figure}

In cross-corpus evaluation,
a model is presented with data from an unknown source.
To simulate such a situation,
we train a model on a single dataset.
For our experiment,
we choose {\agender} as it is the only dataset
that contains a considerable amount of child speech.
In \cref{fig:cross-corpus}, 
we report the performance of this model as
\emph{cross-corpus} 
on the remaining datasets,
namely {\commonvoice},
{\timit},
and {\voxceleb}.
For comparison,
we also include \emph{in-domain} results
by the model trained on all datasets.

Removing in-domain data from the training
generally leads to a performance drop on all datasets.
On {\commonvoice}, the effect is yet small:
plus one year for age 
and minus one percentage point for gender.
Whereas on {\timit} and {\voxceleb} it is quite significant:
\ac{MAE} increases by four to five years for age 
and \ac{ACC} decreases by more than ten percentage points for gender.
In the lower part of \cref{fig:cross-corpus},
we visualise gender and age predictions 
aggregated over the three datasets.
The confusion matrices reveal that the cross-domain model
has problems in predicting females,
while the distribution plots show that age is generally underestimated.

We conclude more and diverse data is needed
to build a robust age and gender model, 
especially when using an upsampled $8$\,kHz dataset as training.

\subsection{Varying the number of layers}
\label{sec:layers}

\begin{figure}[ht]
    \centering
    \includegraphics[width=\columnwidth]{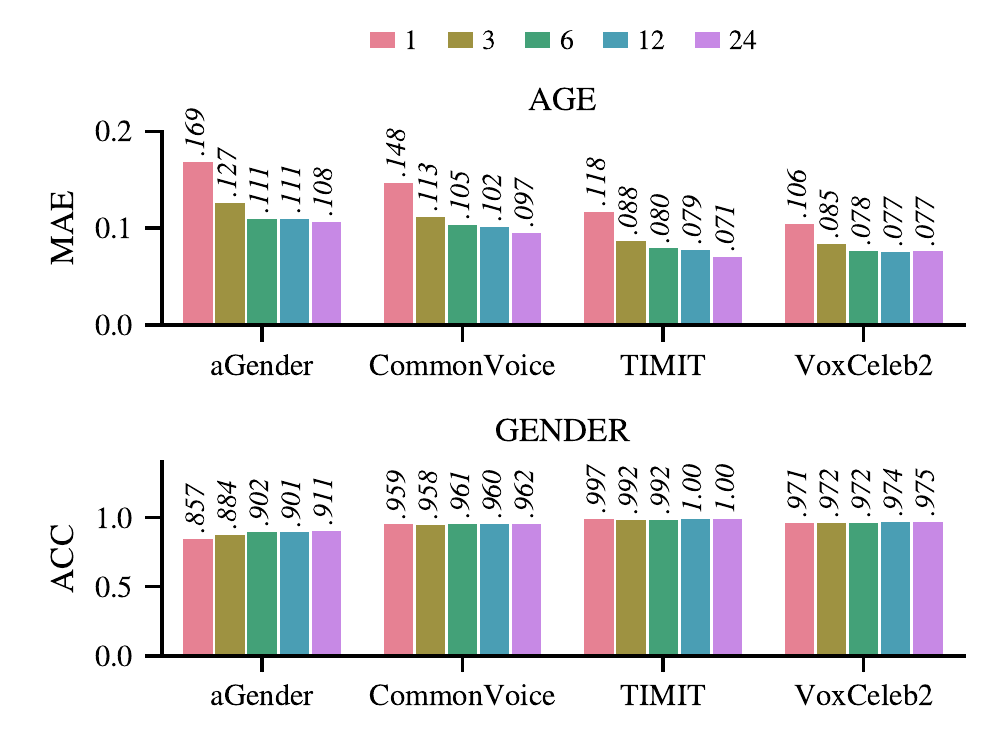}
    \caption{Results with different number of transformer layers. For gender, a single transformer layer seems sufficient, whereas for age, six layers provide a good trade-off between speed and accuracy  (age: MAE in years/100).}
    \label{fig:layers}
\end{figure}

In the experiments reported so far,
we have used all 24 transformer layers.
Dropping some of top layers reduces the footprint of a model.
However,
too few layers may degrade the ability of the model to properly learn a task.
To investigate the effect of reducing transformer layers,
we run experiments with a varying number of layers.
As we see in \cref{fig:layers},
results generally improve with more layers,
though the effect is smaller
for gender than for age.
In fact,
a single transformer layer seems sufficient
to adequately model gender.
For age,
we can observe a 
considerable
performance drop 
for less than six layers,
while with more than six layers,
there is only a marginal increase.

We conclude that using
six transformer layers provides
a good trade-off between accuracy and speed.
This reduces the number of parameters by a factor of $3.5$ to $90.8$M
and inference time by a factor of to $3$ to $12.4 \pm 3.3$ms.

\subsection{Comparison classical modelling approach}

\begin{table}
    \footnotesize{
        \begin{tabular}{lcccl}
            \toprule
            {} &    \bf{Age} & \bf{Gender} & \bf{Combined} & \bf{Training} \\
            \bf{System} & 4-class & 3-class  & 7-class & \\
            \midrule
            \multicolumn{5}{c}{\emph{Development}} \\
            \midrule
            baseline \cite{kockmann10b_interspeech} & .56 / .55 &  .82 / .87 & .54 / .54 & {\agender} \\
            {\wtov} & .60 / .60 & .84 / \textbf{.92} & .58 / .60 & {\agender} \\
            {\wtov} & \textbf{.61} / \textbf{.61} & \textbf{.86} / .91 & \textbf{.59} / \textbf{.61} & All \\
            \midrule
            \multicolumn{5}{c}{\emph{Test}} \\
            \midrule
            baseline \cite{kockmann10b_interspeech} & .52 / .51 &  .83 / .86 & \emph{N/A} & {\agender} \\
            {\wtov} & .60 / \textbf{.57} & .86 / \textbf{.90} & \textbf{.57} / \textbf{.56} & {\agender} \\
            {\wtov} & \textbf{.61} / \textbf{.57} & \textbf{.87} / .88 & \textbf{.57} / \textbf{.56} & All \\
            \midrule
            \bottomrule
        \end{tabular}
    }
    \caption{Comparison with baseline system based on handcrafted features. Results are reported in terms of \acs{UAR} / \acs{ACC} on the development set (upper part) and test set (lower part). In case of age, we map the predictions of our model to the four classes child, youth, adult, and senior. In the combined task, age and gender are jointly represented by seven classes.}    
    \label{tab:baseline}    
\end{table}

Finally,
we compare performance
to a classic modelling approach
based on hand-crafted features.
As a baseline, we choose the winner system of the
2010 Interspeech Paralinguistic Challenge \cite{Schuller2010}.
It implements a combination of
Gaussian Mixture Models (GMM) and
Support Vector Machines (SVM), 
followed by linear Gaussian
backends and logistic regression-based fusion,
which uses as input a large feature set of 
acoustic, prosodic, and voice quality features.
For more information see \citet{kockmann10b_interspeech}.

In \cref{tab:baseline}, 
we compare the performance of the baseline system
with our best performing model
(24 layers)
when trained either only on {\agender}
or on all datasets introduced in \cref{sec:database}.
In case of age,
we map the continuous predictions of our model
to the four classes child, youth, adulate, and senior
as proposed by the challenge organisers.
In addition,
we include results 
for the combined age/gender task
with seven classes used in the challenge \cite{Schuller2010}.
Depending on the task,
\ac{UAR} and \ac{ACC} 
of the baseline is improved
by 4--10 percentage points.
Using all datasets during training
provides an additional,
yet small boost.

We can conclude that deep learning 
improves the accuracy compared to 
a classic modelling based on
manual feature engineering.

\subsection{Age/gender prediction of emotional data}
 When testing the published model on the Berlin emotional database \cite{emodb}, the MAE for all samples age prediction is 8.35 and the UAR for binary gender prediction 96.04. When only the neutral samples are used, the MAE drops to 5.94
and the UAR to 100. Clearly, the acted emotional expression jeopardizes the quality of our model, as it has been trained on non-emotional data.



\section{Conclusions and Outlook}
\label{sec:conclusion}

We performed experiments on age and gender prediction based on four datasets and a fine tuned transformer architecture.
A model trained on all data sets, together with the test, train, and develpment splits, has been made public and can be used as a baseline for other authors.
We will continue to investigate age detection by using other model architectures and perhaps combining them with expert features.
Especially speech data from children is sparse and we will look for such data or try to synthesise data with generative models.


\section{Acknowledgements}
This research has been partly funded by the European EASIER (Intelligent Automatic Sign Language Translation) project (Grant Agreement number: 101016982) as well as the European MARVEL (Multimodal Extreme Scale Data Analytics for Smart Cities Environments) project (Grant Agreement ID: 975337).



\section{\refname}

\printbibliography[heading=none]

@inproceedings{Eyben2010,
author = {Eyben, F. and Wöllmer, M. and Schuller, B.},
booktitle = {Proceedings of the 18$^{th}$ ACM international conference on Multimedia},
doi = {},
isbn = {978-1-60558-933-6},
issn = {},
title = {{openSMILE –- the Munich versatile and fast open-source audio feature extractor}},
volume = {},
year = {2010},
pages = {1459--1462}
}

@inproceedings{Schuller2016,
author = {Schuller, B. and Steidl, S. and Batliner, A. and Hirschberg, J.and Burgoon, J.K. and Baird, A. and Elkins, A. and Zhang1, Y. and Coutinho, E. and Evanini, K.},
booktitle = {Proceedings of the 17$^{th}$ Annual Conference of the International Speech Communication Association, INTERSPEECH 2016},
doi = {},
isbn = {},
issn = {},
title = {{The INTERSPEECH 2016 Computational Paralinguistics Challenge: Deception, Sincerity \& Native Language}},
volume = {},
year = {2016}
}

@inproceedings{Metze2007,
author = {Metze, F. and Ajmera, J. and Englert, R. and Bub, U. and Burkhardt, F. and Stegmann, J. and M{\"{u}}ller, C. and Huber, R. and Andrassy, B. and Bauer, J.G. and Littel, B.},
booktitle = {ICASSP, IEEE International Conference on Acoustics, Speech and Signal Processing - Proceedings},
volume = {4},
year = {2007}
}

@inproceedings{Bocklet2008,
author = {Bocklet, T. and Maier, A. and Bauer, J.G. and Burkhardt, F. and N{\"{o}}th, E.},
booktitle = {ICASSP, IEEE International Conference on Acoustics, Speech and Signal Processing -- Proceedings},
title = {{Age and gender recognition for telephone applications based on GMM supervectors and support vector machines}},
year = {2008}
}

@inproceedings{Burkhardt2010,
author = {Burkhardt, F. and Eckert, M. and Johannsen, W. and Stegmann, J.},
booktitle = {Proceedings of the 7$^{th}$ International Conference on Language Resources and Evaluation, LREC 2010},
title = {{A database of age and gender annotated telephone speech}},
year = {2010}
}

@inproceedings{Feld2010,
author = {Feld, M. and Burkhardt, F. and M{\"{u}}ller, C.},
booktitle = {Proceedings of the 11th Annual Conference of the International Speech Communication Association, INTERSPEECH 2010},
title = {{Automatic speaker age and gender recognition in the car for tailoring dialog and mobile services}},
year = {2010}
}

@book{Brue2011,
 title = {Altersbedingte Ver{\"{a}}nderungen der Stimme und Sprechweise von Frauen},
 series = {M{\"{u}}ndliche Kommunikation},
 volume = {7},
 publisher = {Logos Verlag},
 year = {2011},
 author = {Br{\"{u}}ckl, M.},
 address = {Berlin}
}

@inproceedings{Shor2020,
  author={Joel Shor and Aren Jansen and Ronnie Maor and Oran Lang and Omry Tuval and Félix de Chaumont Quitry and Marco Tagliasacchi and Ira Shavitt and Dotan Emanuel and Yinnon Haviv},
  title={{Towards Learning a Universal Non-Semantic Representation of Speech}},
  year=2020,
  booktitle={Proc. Interspeech 2020},
  pages={140--144},
}

@inproceedings{Lingenfelser2010,
author = {Lingenfelser, Florian and Wagner, Johannes and Vogt, Thurid and Kim, Jonghwa and Andr{\'{e}}, Elisabeth},
booktitle = {Proceedings of the 11th Annual Conference of the International Speech Communication Association, INTERSPEECH 2010},
title = {{Age and gender classification from speech using decision level fusion and ensemble based techniques}},
year = {2010}
}

@article{Chawla2002,
author = {Chawla, Nitesh V. and Bowyer, Kevin W. and Hall, Lawrence O. and Kegelmeyer, W. Philip},
journal = {Journal of Artificial Intelligence Research},
title = {{SMOTE: Synthetic minority over-sampling technique}},
year = {2002}
}

@article{Eyben2016,
journal = {IEEE Transactions on Affective Computing},
author = {Eyben, Florian and Scherer, Klaus R. and Schuller, Bjorn W. and Sundberg, Johan and Andre, Elisabeth and Busso, Carlos and Devillers, Laurence Y. and Epps, Julien and Laukka, Petri and Narayanan, Shrikanth S. and Truong, Khiet P.},
title = {{The Geneva Minimalistic Acoustic Parameter Set (GeMAPS) for Voice Research and Affective Computing}},
year = {2016}
}

@Manual{Brue2018,
    title = {irrNA: Coefficients of Interrater Reliability -- Generalized for Randomly Incomplete Datasets},
    author = {Br\"{u}ckl, M and Heuer, F.},
    year = {2018},
    note = {R package version 0.1.4},
    url = {https://CRAN.R-project.org/package=irrNA},
  }

@inproceedings{Katerenchuk2017,
author = {Katerenchuk, Denys},
booktitle = {15th Conference of the European Chapter of the Association for Computational Linguistics, EACL 2017 - Proceedings of Conference},
title = {{Age group classification with speech and metadata multimodality fusion}},
year = {2017}
}

@inproceedings{oates2019robust,
  title={Robust Speech Emotion Recognition Under Different Encoding Conditions.},
  author={Oates, Christopher and Triantafyllopoulos, Andreas and Steiner, Ingmar and Schuller, Bj{\"o}rn W},
  booktitle={INTERSPEECH},
  pages={3935--3939},
  year={2019}
}

@article{sanchez_2022,
   author = {Héctor A Sánchez-Hevia and Roberto Gil-Pita and · Manuel Utrilla-Manso and Manuel Rosa-Zurera and Manuel Utrilla-Manso},
   journal = {Multimedia Tools and Applications},
   title = {Age group classification and gender recognition from speech with temporal convolutional neural networks},
   year = {2022},
}

@INPROCEEDINGS{hechmi_2021,
  author={Hechmi, Khaled and Trong, Trung Ngo and Hautamäki, Ville and Kinnunen, Tomi},
  booktitle={2021 IEEE Automatic Speech Recognition and Understanding Workshop (ASRU)}, 
  title={Voxceleb Enrichment for Age and Gender Recognition}, 
  year={2021},
  volume={},
  number={},
  pages={687-693},
}

@article{Gupta2022,
   author = {Tarun Gupta and Duc Tuan Truong and Tran The Anh and Chng Eng Siong},
   journal = {Proceedings of the Annual Conference of the International Speech Communication Association, INTERSPEECH},
   month = {3},
   pages = {1978-1982},
   publisher = {International Speech Communication Association},
   title = {Estimation of speaker age and height from speech signal using bi-encoder transformer mixture model},
   volume = {2022-September},
   year = {2022},
}

@article{Sadjadi2016,
   author = {Seyed Omid Sadjadi and Sriram Ganapathy and Jason W. Pelecanos},
   journal = {ICASSP, IEEE International Conference on Acoustics, Speech and Signal Processing - Proceedings},
   month = {5},
   pages = {5040-5044},
   publisher = {Institute of Electrical and Electronics Engineers Inc.},
   title = {Speaker age estimation on conversational telephone speech using senone posterior based i-vectors},
   volume = {2016-May},
   year = {2016},
}

@article{Zazo2018,
   author = {Ruben Zazo and Phani Sankar Nidadavolu and Nanxin Chen and Joaquin Gonzalez-Rodriguez and Najim Dehak},
   journal = {IEEE Access},
   month = {3},
   pages = {22524-22530},
   publisher = {Institute of Electrical and Electronics Engineers Inc.},
   title = {Age Estimation in Short Speech Utterances Based on LSTM Recurrent Neural Networks},
   volume = {6},
   year = {2018},
}

@inproceedings{burkhardt-etal-2010-database,
    title = "A Database of Age and Gender Annotated Telephone Speech",
    author = "Burkhardt, Felix  and
      Eckert, Martin  and
      Johannsen, Wiebke  and
      Stegmann, Joachim",
    booktitle = "Proceedings of the Seventh International Conference on Language Resources and Evaluation ({LREC}'10)",
    month = may,
    year = "2010",
    address = "Valletta, Malta",
    publisher = "European Language Resources Association (ELRA)",
}

@misc{timit,
  author = {Garofolo, J. S. and Lamel, L. F. and Fisher, W. M. and Fiscus, J. G. and Pallett, D. S. and Dahlgren, N. L.},
  publisher = {NIST},
  title = {DARPA TIMIT Acoustic Phonetic Continuous Speech Corpus CDROM},
  year = 1993
}

@article{Schuller2010,
   author = {Björn Schuller and Stefan Steidl and Anton Batliner and Felix Burkhardt and Laurence Devillers and Christian Müller and Shrikanth Narayanan},
   journal = {Proceedings of the 11th Annual Conference of the International Speech Communication Association, INTERSPEECH 2010},
   pages = {2794-2797},
   publisher = {International Speech Communication Association},
   title = {The INTERSPEECH 2010 paralinguistic challenge},
   year = {2010},
}

@inproceedings{kockmann10b_interspeech,
   author = "Marcel Kockmann and Luk\'{a}\v{s} Burget and Jan \v{C}ernock\'{y}",
   title = "Brno University of Technology System for Interspeech 2010 Paralinguistic Challenge",
   pages = "2822--2825",
   booktitle = "Proceedings of the 11th Annual Conference of the International Speech Communication Association (INTERSPEECH 2010)",
   journal = "Proceedings of Interspeech - on-line",
   volume = 2010,
   number = 9,
   year = 2010,
   location = "Makuhari, Chiba, JP",
   publisher = "International Speech Communication Association",
}

@article{wagner2023dawn,
  title={Dawn of the Transformer Era in Speech Emotion Recognition: Closing the Valence Gap},
  author={Wagner, Johannes and Triantafyllopoulos, Andreas and Wierstorf, Hagen and Schmitt, Maximilian and Burkhardt, Felix and Eyben, Florian and Schuller, Bj{\"o}rn W},
  journal={IEEE Transactions on Pattern Analysis and Machine Intelligence},
  pages={1--13},
  year={2023},
}

@inproceedings{baevski2020wav2vec,
  author={Baevski, Alexei and Zhou, Yuhao and Mohamed, Abdelrahman and Auli, Michael},
  title={wav2vec 2.0: A framework for self-supervised learning of speech representations},
  booktitle={Advances in Neural Information Processing Systems (NeurIPS)},
  year={2020},
  pages={12449--12460},
  address={Vancouver, BC, Canada},
}

@article{hsu2021robust,
  title={Robust wav2vec 2.0: Analyzing domain shift in self-supervised pre-training},
  author={Hsu, Wei-Ning and Sriram, Anuroop and Baevski, Alexei and Likhomanenko, Tatiana and Xu, Qiantong and Pratap, Vineel and Kahn, Jacob and Lee, Ann and Collobert, Ronan and Synnaeve, Gabriel and Auli, Michael},
  journal={arXiv preprint arXiv:2104.01027},
  year={2021}
}

@article{Nagrani2020,
   author = {Arsha Nagrani and Joon Son Chung and Weidi Xie and Andrew Zisserman},
   journal = {Computer Speech \& Language},
   month = {3},
   pages = {101027},
   publisher = {Academic Press},
   title = {Voxceleb: Large-scale speaker verification in the wild},
   volume = {60},
   year = {2020},
}

@article{Ardila2020,
   author = {Rosana Ardila and Megan Branson and Kelly Davis and Michael Henretty and Michael Kohler and Josh Meyer and Reuben Morais and Lindsay Saunders and Francis M. Tyers and Gregor Weber},
   journal = {LREC 2020 - 12th International Conference on Language Resources and Evaluation, Conference Proceedings},
   pages = {4218-4222},
   publisher = {European Language Resources Association (ELRA)},
   title = {Common voice: A massively-multilingual speech corpus},
   year = {2020},
}

@article{Levitan2016,
   author = {Sarah Ita Levitan and Taniya Mishra and Srinivas Bangalore},
   journal = {Proceedings of the International Conference on Speech Prosody},
   pages = {84-88},
   publisher = {International Speech Communication Association},
   title = {Automatic identification of gender from speech},
   volume = {2016-January},
   year = {2016},
}

@article{schoetz_2007,
   author = {Susanne Schötz},
   journal = {Lecture Notes in Computer Science (including subseries Lecture Notes in Artificial Intelligence and Lecture Notes in Bioinformatics)},
   pages = {88-107},
   publisher = {Springer Verlag},
   title = {Acoustic analysis of adult speaker age},
   volume = {4343 LNAI},
   year = {2007},
}

@article{tursunov2021age,
  title={Age and gender recognition using a convolutional neural network with a specially designed multi-attention module through speech spectrograms},
  author={Tursunov, Anvarjon and Choeh, Joon Yeon and Kwon, Soonil},
  journal={Sensors},
  volume={21},
  number={17},
  pages={5892},
  year={2021},
  publisher={MDPI}
}

@article{alnuaim2022speaker,
  title={Speaker gender recognition based on deep neural networks and ResNet50},
  author={Alnuaim, Abeer Ali and Zakariah, Mohammed and Shashidhar, Chitra and Hatamleh, Wesam Atef and Tarazi, Hussam and Shukla, Prashant Kumar and Ratna, Rajnish},
  journal={Wireless Communications and Mobile Computing},
  volume={2022},
  pages={1--13},
  year={2022},
  publisher={Hindawi Limited}
}

@article{kwasny2020joint,
  title={Joint gender and age estimation based on speech signals using x-vectors and transfer learning},
  author={Kwasny, Damian and Hemmerling, Daria},
  journal={arXiv preprint arXiv:2012.01551},
  year={2020}
}

@inproceedings{wolf-etal-2020-transformers,
    title = "Transformers: State-of-the-Art Natural Language Processing",
    author = "Wolf, Thomas  and
      Debut, Lysandre  and
      Sanh, Victor  and
      Chaumond, Julien  and
      Delangue, Clement  and
      Moi, Anthony  and
      Cistac, Pierric  and
      Rault, Tim  and
      Louf, Remi  and
      Funtowicz, Morgan  and
      Davison, Joe  and
      Shleifer, Sam  and
      von Platen, Patrick  and
      Ma, Clara  and
      Jernite, Yacine  and
      Plu, Julien  and
      Xu, Canwen  and
      Le Scao, Teven  and
      Gugger, Sylvain  and
      Drame, Mariama  and
      Lhoest, Quentin  and
      Rush, Alexander",
    booktitle = "Proceedings of the 2020 Conference on Empirical Methods in Natural Language Processing: System Demonstrations",
    month = oct,
    year = "2020",
    address = "Online",
    publisher = "Association for Computational Linguistics",
    pages = "38--45",
}

@inproceedings{emodb,
    author = {Burkhardt, Felix and Paeschke, Astrid and Rolfes, M. and Sendlmeier, Walter and Weiss, Benjamin},
    year = {2005},
    month = {09},
    pages = {1517-1520},
    title = {A database of German emotional speech},
    volume = {5},
    booktitle = {9th European Conference on Speech Communication and Technology},
    doi = {10.21437/Interspeech.2005-446}
}

\end{document}